\documentclass[cits]{PoS}

\newcommand{\ptee}{$ \it{p}_{\textrm{T,ee}}\ $}
\newcommand{\pte}{$ \it{p}_{\textrm{T,e}}\ $}

\newcommand{\gevc}{GeV/\textit{c}}
\newcommand{\gevcs}{GeV/\textit{c}$ ^2$}

\usepackage{sansmath}
\usepackage [ english ]{ babel }
\usepackage {lmodern}
\usepackage [T1]{fontenc}
\usepackage[pdftex]{graphicx}

\usepackage{amsmath} 
\usepackage{amssymb}
\usepackage{mathptmx}
\usepackage{times}
\usepackage{subfigure}
\usepackage[utf8]{inputenc}

\usepackage[capitalize]{cleveref}  
\usepackage{cite}

\title{Dielectron production at low transverse momentum in Pb--Pb collisions at $ \sqrt{s_{\mathrm{NN}}}=5.02 $ TeV with ALICE}
\author{\speaker {Sebastian Lehner}\\
	on behalf of the ALICE collaboration\\
	\\
	Stefan-Meyer Institute for Subatomic Physics (Austrian Acadamey of Sciences), Austria\\
	E-mail:	\email{selehner@cern.ch}}
\abstract{
Dielectrons probe a wide range of phenomena in heavy-ion collisions. These include light- and heavy-flavour meson production, thermal radiation and coherent photo-production. The latter process is distinguished by dielectron production at low transverse-pair momentum ($ \it{p}_{\textrm{T,ee}}$). Transverse momentum spectra of dielectrons in central and peripheral Pb--Pb collisions are extracted and compared to the corresponding expectations. In central collisions the data fit the expected spectrum, which is dominated by semi-leptonic decays of correlated heavy-flavour hadrons. In peripheral collisions, the data exhibit an excess at low \ptee with respect to hadronic and thermal sources. The observed excess yield is compatible with calculations for dielectron production from coherent photon-photon interactions.
}
\FullConference{The Seventh Annual Large Hadron Collider Physics (LHCP2019)}
\ShortTitle{Dielectron production at low transverse momentum in Pb--Pb collisions}

\begin{document}
\setlength{\parindent}{0ex}
\section{Introduction}
Dielectrons (e$^+$e$^-$) are produced at every stage of a heavy-ion collision. Therefore, their analysis offers insights into a large variety of related questions. Information from the earliest stages is well conserved since leptons undergo practically no final state interaction in the strongly interacting medium.\\
Besides dielectron production due to hadronic interactions of the colliding projectiles, electromagnetic interactions are expected to result in coherent photo-production of dielectrons. This process is typically studied in ultra-peripheral collisions (UPC), in which no hadronic interaction between the colliding nuclei takes place. The coherence condition asserts that the interacting electromagnetic fields probe the whole nucleus. For UPC the assumption of coherence is less problematic than in collisions with hadronic interactions since the impact parameter is large compared to the size of the nuclei and the nuclei are not broken up. Theoretically such processes are treated in the Equivalent Photon Approximation (EPA), where a photon flux is associated to the electromagnetic fields of the nuclei. In dielectron production from coherent photon-photon interactions, these photons scatter into e$^+$e$^-$ pairs. Corresponding calculations indicate that these dielectrons should primarily be produced at low transverse momemtum ($\it{p}_{\textrm{T,ee}}<0.2$ \gevc) over a wide invariant mass range \cite{mariola, coh, klein}.\\
A closely related photon-induced process is coherent photo-nuclear production of vector mesons and their subsequent decay into dielectrons. In this process one nucleus emits a photon which interacts with the other nucleus and produces a vector meson. Photo-nuclear production is considered to be the most likely source of the $ J/\psi $ excess observed by \mbox{ALICE} \cite{alicejpsi} and STAR \cite{starjpsi} in peripheral collisions.\\
Indications for dilepton production via photon-photon interactions in hadronic heavy-ion collisions were reported at RHIC and the LHC. The STAR collaboration found an excess with respect to the hadronic expectation of low-$ \it{p}_{\textrm{T,ee}}$ dielectrons in Au--Au and U--U collisions, which is compatible with calculations for coherent photo-production \cite{star,mariola, coh, klein} and which cannot be linked to hadronic sources \cite{phsd}. The ATLAS collaboration observes photo-produced dimuons in Pb--Pb collisions, which exhibit a centrality dependent broadening effect of their angular pair correlations that seems not to be present in UPC \cite{atlasdimu, kleinaco}. In the following, the first dedicated measurement of low-mass dielectron production at low \ptee in Pb--Pb collisions at $ \sqrt{s_{\mathrm{NN}}}=5.02 $ TeV is reported and compared to various models of dielectron photo-production.


\section{Data Analysis and Results}

To isolate photon-photon interactions from photon-nuclear interactions, this analysis is restricted to the invariant mass ($ \it{m}_{\textrm{ee}}$) range of 1.1--2.7 \gevcs, which does not contain vector resonances. Among the known hadronic dielectron sources, typically called the cocktail, the  semi-leptonic decays of correlated charm and beauty hadrons are the only relevant ones in this mass range. The data, which were recorded in 2015, are compared to dielectron yields corresponding to these decays, which are based on the binary scaled $ \mathrm{\text{c}\overline{\text{c}}}$ and $ \mathrm{\text{b}\overline{\text{b}}}$ production cross sections ($\sigma_\mathrm{\text{c}\overline{\text{c}},\text{b}\overline{\text{b}}}$) in pp collisions. The rapidity ($y$) differential charm cross section of $ \mathrm{d\sigma_{\text{c}\overline{\text{c}}}/d}y|_{y=0}=0.792^{+ 0.137}_{-0.101}$ mb is obtained from a FONLL \cite{FONLL} extrapolation to $ \sqrt{s}=5.02 $ TeV of ALICE measurements at $ \sqrt{s}=7 $ TeV \cite{ccbar}. For beauty production the cross-section of $ \sigma_{\text{b}\overline{\text{b}}}=0.204^{+ 0.035}_{-0.034}$ mb is obtained from a FONLL extrapolation to $ \sqrt{s}=5.02 $ TeV of LHCb measurements at $ \sqrt{s}=7 $ TeV \cite{bbar}. Hadronisation and hadron decays are simulated using PYTHIA 6.4 \cite{pythia}.  In addition to the cocktail, calculations for dielectron production from thermal radiation and photon-photon interaction are included in the comparison to data.\\
Electron and positron tracks are selected using the Particle Identification (PID) information from the Inner Tracking System and the Time Projection Chamber. For tracks with an associated hit in the Time-Of-Flight (TOF) detector, an additional selection requirement is applied on the TOF information \cite{alice,aliceper}.\\
In the analysis of dielectrons the main background comes from combinatorial dielectrons, i.e. pairs of an electron and positron track which do not have the same microscopic particle origin and which were not produced in decays of correlated heavy-flavour hadrons. Simulations show that the majority of these pairs contain tracks that originate from photon-conversion processes, i.e. the conversion of a real photon into a dielectron in the material surrounding the interaction point. To suppress these tracks a Boosted Decision Tree (BDT) classifier is trained on Monte Carlo (MC) simulation data to learn the characteristics of conversion tracks in terms of track quality parameters and kinematics. Applying selection criteria on the output of this classifier on real data suppresses combinatorial dielectrons with a conversion leg.
\begin{figure}[h]
	\centering
	\includegraphics[scale= 0.24]{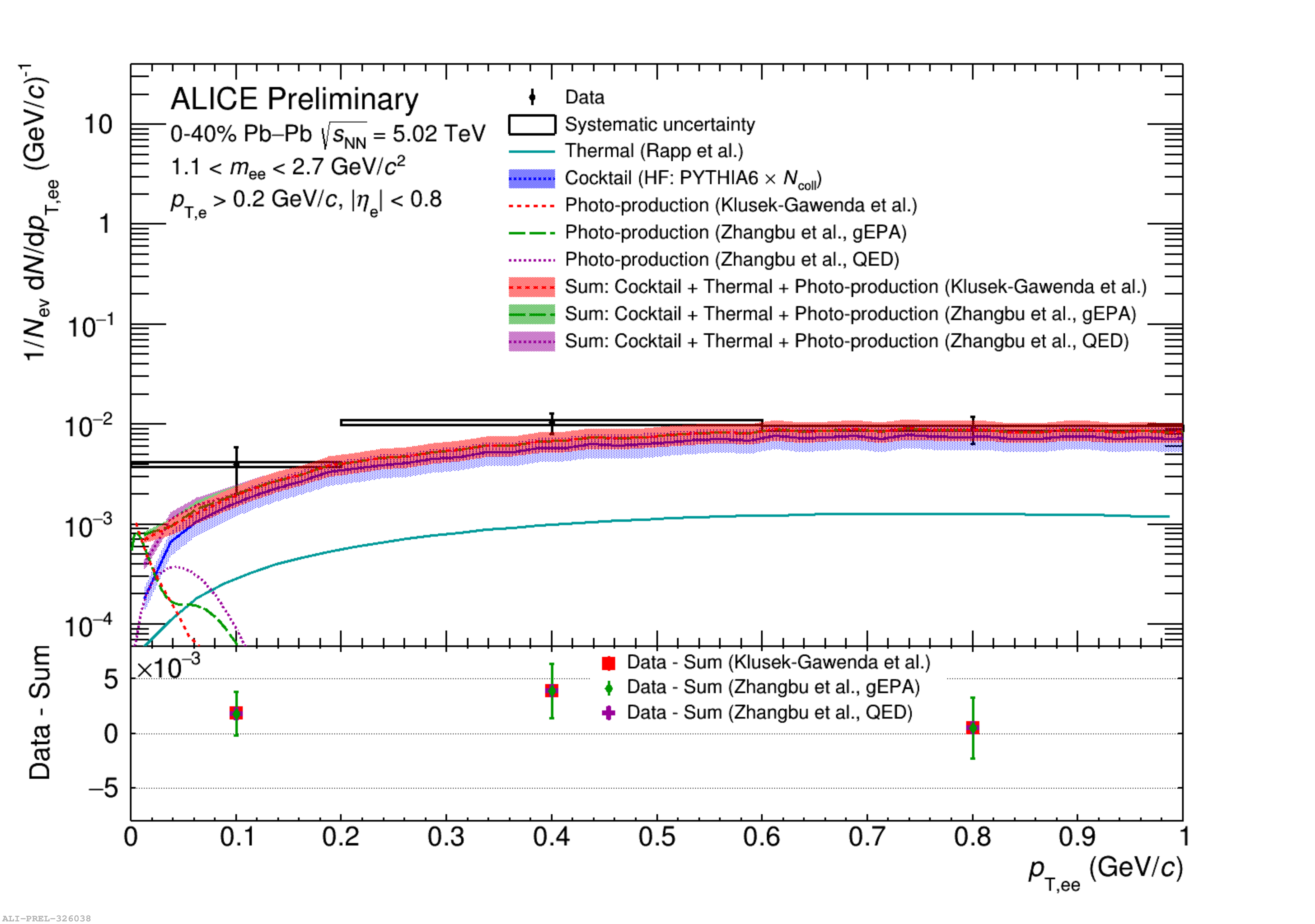}	
	\caption{Transverse momentum spectrum of dielectrons in the centrality interval of 0--40\% compared to the cocktail, along with models of thermal radiation and photo-production.}	
	\label{res0040}
\end{figure} 
\begin{figure}[h]
	\centering
	\includegraphics[scale= 0.24]{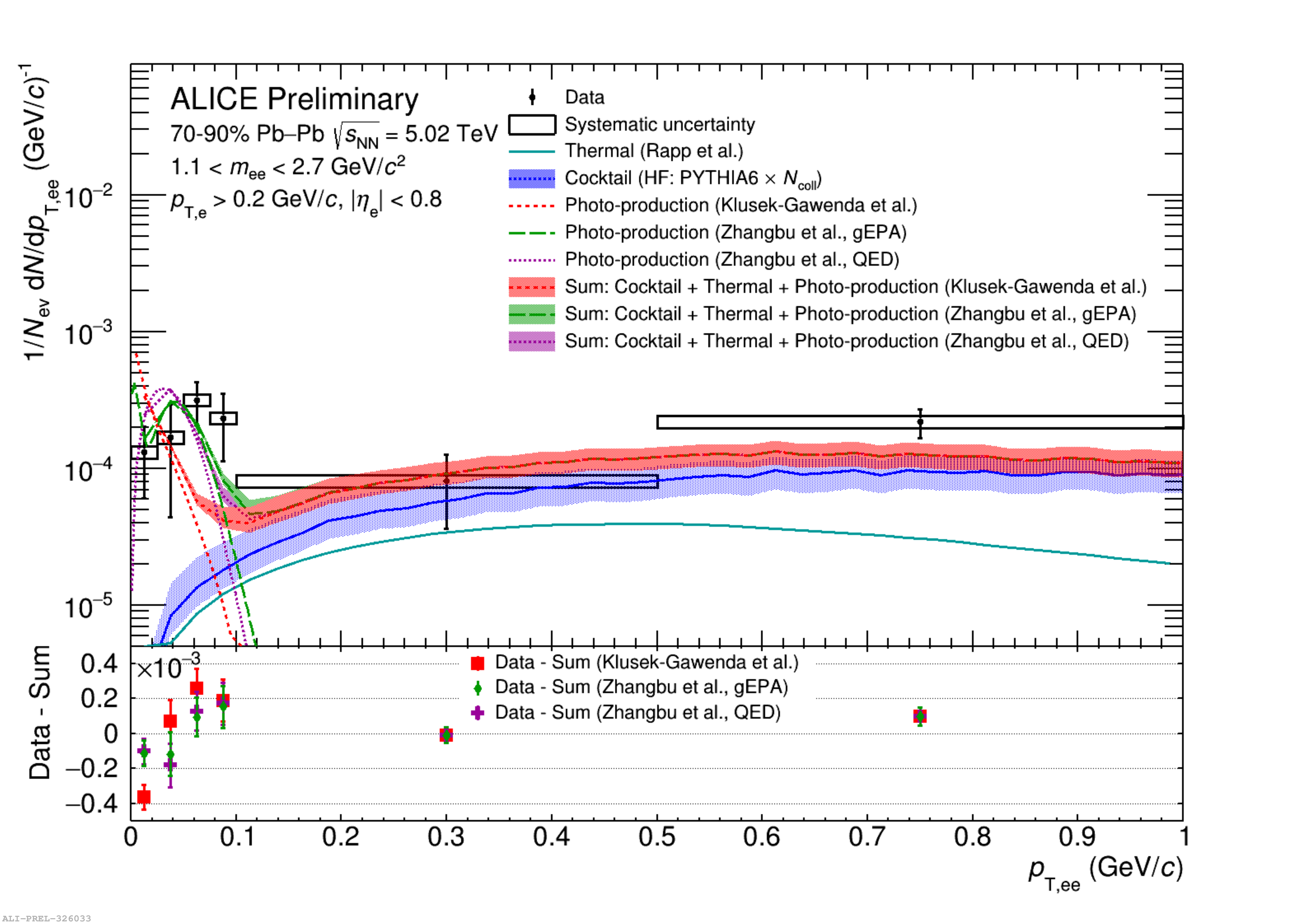}		
	\caption{Transverse momentum spectrum of dielectrons in the centrality interval of 70--90\% compared to the cocktail, along with models of thermal radiation and photo-production.}
	\label{res7090}
\end{figure}
\\
The combinatorial background from the remaining tracks is subtracted using the like-sign subtraction method \cite{alberto}. The dielectron yield in the kinematic range of single-electron \pte$>0.2$ \gevc\ and single-electron pseudorapidity $|\eta_\textrm{e}|<0.8$ is obtained from the corresponding measured dielectron yields by applying $ \it{m}_{\textrm{ee}}$, $ \it{p}_{\textrm{T,ee}}$, and centrality dependent correction factors, which are based on MC simulations.\\
The analysis is carried out in the centrality classes 0--40\% and 70--90\%. Centralities are estimated using the V0 detectors \cite{alicecent}. Dielectron production from hadronic and thermal sources rises more quickly with centrality  than photo-production, which shows only a modest centrality dependence \cite{mariola}. Therefore, the centrality class 70--90\% is more sensitive to photo-production than 0--40\%.\\
Variations of the PID selection criteria showed no significant signal contribution from hadron contamination. The systematic uncertainties of the spectra are estimated from their variation under different track quality, PID selection, and conversion rejection criteria.\\ 
\Cref{res0040,res7090} show the dielectron \ptee spectra in Pb--Pb collisions in the 0--40\% and 70--90\% centrality classes. The data are compared to the sum of the cocktail, a model of thermal radiation by Rapp et al.\ \cite{rapp,rapppriv} and models of photo-production. Three different photo-production models are compared. The model by Kłusek-Gawenda et al.\ uses EPA and Woods-Saxon nuclear form factors \cite{mariola,marpriv}. Zhangbu et al.\ provide calculations using a generalized EPA (gEPA) approach which results in an impact parameter dependent \ptee shape of the spectra \cite{broad, zhapriv}. In addition, Zhangbu et~al.\ carried out a full leading order QED calculation \cite{zhapriv}. Effects of the finite tracking resolution on, e.g.\ the single-leg \pte are not implemented in these models. These effects are, however, expected to play a minor role in the comparison to data.\\
In the 40\% most central collisions the spectrum is in agreement with the cocktail (see \cref{res0040}). Contributions from thermal radiation and photo-production cannot be identified given the present precision of the measured spectrum.\\
The peripheral data set of 70--90\% centrality shows an excess of dielectrons with respect to the cocktail at \ptee$<0.1$ \gevc\ (see \cref{res7090}). Both photo-production calculations by Zhangbu et~al.\ are compatible with the data. The data exhibit a peak at slightly higher \ptee than these models. For the model by Kłusek-Gawenda et al.\ the peak is located at even lower \ptee and thus the discrepancy to the data is more pronounced. The overall yield at \ptee$<0.1$ \gevc\ is described well by all three photo-production models.
 
\section{Conclusion and Outlook}
The presented spectrum in peripheral collisions can be regarded as an indication of dielectron production by coherent photon-photon interactions in collisions with hadronic overlap. The measured excess yield with respect to the cocktail and thermal sources is well described by photo-production models. A significant difference is found in the calculated shapes of the \ptee\ spectra among the compared models. While tracking resolution effects are not yet considered in the comparison, it seems the calculations by Zhangbu et al.\ are favoured by the data.\\
In central collisions the data are in line with the cocktail, which is expected to be the only relevant dielectron source in this centrality class and mass range.\\
Apart from including resolution effects in the model calculations, an extension of the analysis in terms of mass and centrality range is foreseen, which should allow for more differential studies. These may address the geometry of the initial electromagnetic field and its fluctuations, effects of the strong magnetic field, and interactions with the medium \cite{broad,bfield,fluc, kleinaco}. Data recorded in 2018 are currently becoming available and will approximately double the data set size in peripheral collision, which will further benefit these analyses.   

\acknowledgments
We are indebted to Mariola Kłusek-Gawenda, Xu Zhangbu, Ralf Rapp and their collaborators for providing the shown calculations from their models of photon-photon and thermal dielectron production. 
\bibliographystyle{jhep}
\bibliography{lhcp} 

\end{document}